 \documentclass[12pt,apjl]{emulateapj}

\usepackage{lineno}
\usepackage{natbib}
\usepackage{amsmath}
\usepackage{graphicx}
\usepackage{wasysym}
\usepackage{txfonts}
\usepackage{xspace}
\usepackage{url}
\usepackage{textcomp}
\usepackage{multirow}
\usepackage{lipsum}

\newcommand{\ulx}{NGC\,7793~P13\xspace}
\newcommand{\Pdot}{\ensuremath{\dot{P}}\xspace}
\newcommand{\sps}{\ensuremath{\text{s\,s}^{-1}}\xspace}

\newcommand{\swift}{\textsl{Swift}\xspace}

\newcommand{\xmm}{\textsl{XMM-Newton}\xspace}

\newcommand{\nustar}{\textsl{NuSTAR}\xspace}

\newcommand{\msun}{\ensuremath{\text{M}_{\odot}}\xspace}

\newcommand{\ergcms}{\ensuremath{\text{erg\,cm}^{-2}\text{s}^{-1}}\xspace}
\newcommand{\ergps}{\ensuremath{\text{erg\,s}^{-1}}\xspace}
\newcommand{\heii}{\ensuremath{\rm He\,{\small II}}}

\shorttitle{Discovery of pulsations from NGC\,7793~P13}
\shortauthors{F\"urst et al.}

\begin{document}

\title{Discovery of coherent pulsations from the Ultraluminous X-ray Source NGC\,7793 P13}

\author{F.~F\"urst\altaffilmark{1}}
\author{D.~J.~Walton\altaffilmark{2,1}}
\author{F.~A.~Harrison\altaffilmark{1}}
\author{D.~Stern\altaffilmark{2}}
\author{D.~Barret\altaffilmark{3}}
\author{M.~Brightman\altaffilmark{1}}
\author{A.~C.~Fabian\altaffilmark{4}}
\author{B.~Grefenstette\altaffilmark{1}}
\author{K.~K.~Madsen\altaffilmark{1}}
\author{M.~J.~Middleton\altaffilmark{5}}
\author{J.~M.~Miller\altaffilmark{6}}
\author{K.~Pottschmidt\altaffilmark{7,8}}
\author{A.~Ptak\altaffilmark{8}}
\author{V.~Rana\altaffilmark{1}}
\author{N.~Webb\altaffilmark{3}}

\altaffiltext{1}{Cahill Center for Astronomy and Astrophysics, California Institute of Technology, Pasadena, CA 91125, USA; \email{fuerst@caltech.edu}}
\altaffiltext{2}{Jet Propulsion Laboratory, California Institute of Technology, Pasadena, CA 91109, USA}
\altaffiltext{3}{IRAP/CNRS, 9 Av. colonel Roche, BP 44346, F-31028 Toulouse cedex 4, France and
Universit\'e de Toulouse III Paul Sabatier / OMP, Toulouse, France}
\altaffiltext{4}{Institute of Astronomy, Madingley Road, Cambridge CB3 0HA, UK}
\altaffiltext{5}{Department of Physics and Astronomy, University of Southampton, Highfield, Southampton SO17 1BJ, UK}
\altaffiltext{6}{Department of Astronomy, The University of Michigan, Ann Arbor, MI 48109, USA}
\altaffiltext{7}{CRESST, Department of Physics, and Center for Space Science and Technology, UMBC, Baltimore, MD 21250, USA}
\altaffiltext{8}{NASA Goddard Space Flight Center, Greenbelt, MD 20771, USA}

\begin{abstract}

We report the detection of coherent pulsations from the ultraluminous X-ray source
NGC\,7793~P13. The $\approx$0.42\,s nearly sinusoidal pulsations were initially discovered in broadband X-ray observations using  \xmm and
\nustar taken in 2016.  We subsequently also found pulsations in archival \xmm data  
taken in 2013 and 2014. The significant ($\gg5\sigma$) detection of coherent pulsations demonstrates that the compact object in P13 is a neutron star, 
and given the observed peak luminosity of $\approx10^{40}$\,\ergps (assuming isotropy), it is well above the Eddington limit for a 1.4\msun accretor.  This makes P13 the second ultraluminous X-ray source known to be powered by an accreting neutron star. The pulse period varies between epochs, with a slow but persistent spin up over
the 2013--2016 period. This spin-up indicates a magnetic field of $B\approx1.5\times10^{12}$\,G, typical of many Galactic accreting pulsars.   
The most likely explanation for the extreme luminosity is a high degree of beaming, however this is difficult to reconcile with the sinusoidal pulse
profile.

\end{abstract}

.


\section{Introduction}
\label{sec:intro}

Due to their high luminosities ($L >10^{39}\,\ergps$), most ultraluminous X-ray sources (ULXs) have been thought to harbor black holes (BHs)
with masses ranging from  $M\approx10\,\msun$, consistent with a stellar remnant accreting above the Eddington rate (e.g., \citealt{Poutanen07, Middleton15}), to intermediate mass BHs \citep[$M\approx10^{2-5}$\,\msun; e.g.,][]{Miller04} in a sub-Eddington disk accretion regime.  
The discovery of coherent pulsations in the ULX M82~X-2 showed that the compact object in this system is a neutron star \citep{bachetti14a}.
M82~X-2 reaches X-ray luminosities of $2\times10^{40}\,\ergps$,  demonstrating that accreting neutron stars can reach luminosities more than 100 times Eddington (assuming
$M_{\rm{NS}}\approx1.4\,\msun$).

Accreting magnetized  neutron stars can reach these apparent super-Eddington luminosities through a number of mechanisms.     High magnetic fields  collimate the accretion flow, allowing material to accrete onto the polar regions while radiation escapes from the sides of the column \citep{basko76a}.    In addition, large magnetic fields reduce the scattering cross section for electrons \citep{herold79a}, reducing the radiation pressure and increasing the effective Eddington luminosity.     The combination of these effects with the consequent geometric beaming have been used to explain known super-Eddington local sources such as SMC X-1 \citep[e.g.,][]{coe81a}.   

A  very highly magnetized (magnetar-like) neutron star has  been invoked to explain the extreme luminosity of M82~X-2 \citep[e.g.,][]{eksi15a,dallosso15a, mushtukov15a}.  It is difficult, however, to explain the near-sinusoidal pulse profile  in the context of a highly beamed system.  
In contrast, some theoretical work suggests the field in M82~X-2 may be relatively low ($10^9$\,G), based on the ratio of the spin-up rate to the luminosity, which is an order of magnitude lower than typical X-ray pulsars \citep{kluzniak15a}.    These authors argue that a disk truncated at a large radius, as would occur for a high B-field system, would not provide the required lever arm to power the observed spin-up.     The nature of ULX pulsars is very much in question, since no model can explain all the observed characteristics. 

The ULX NGC\,7793 P13 (hereafter P13; \citealt{Read99}) 
is one the few ULXs with a dynamical mass constraint of the compact object and a well classified optical companion \citep[spectral type B9Ia,][]{Motch11}.
Optical monitoring
revealed a $\approx$64\,d photometric period also seen in  the radial velocity of the
\heii\ emission. Adopting this as the orbital period of the binary system, \cite{motch14a}
derive a dynamical mass estimate for the accretor of 3--15\,\msun, assuming a BH. This constraint, together with a peak luminosity of
$L_{\rm X} > 6\times10^{39}$ \ergps, makes P13 a prime example for a super-Eddington system.

Here we report on new \xmm \citep{xmmref} and \nustar \citep{harrison13a} X-ray observations of P13 in which we detect coherent pulsations, requiring
that P13 hosts a highly super-Eddington neutron star accretor.
\footnote{During preparation of this manuscript, \citet[submitted]{israel16a} reported on an independent discovery of this period in archival \xmm data. Our study includes newer \xmm and \nustar data which extend the investigation to higher energies and cover a longer time range.}

\section{Observations and Data Reduction}
\label{sec_red}

\begin{figure*}
\hspace*{-1.0cm}
\hspace{0.6cm}
\plotone{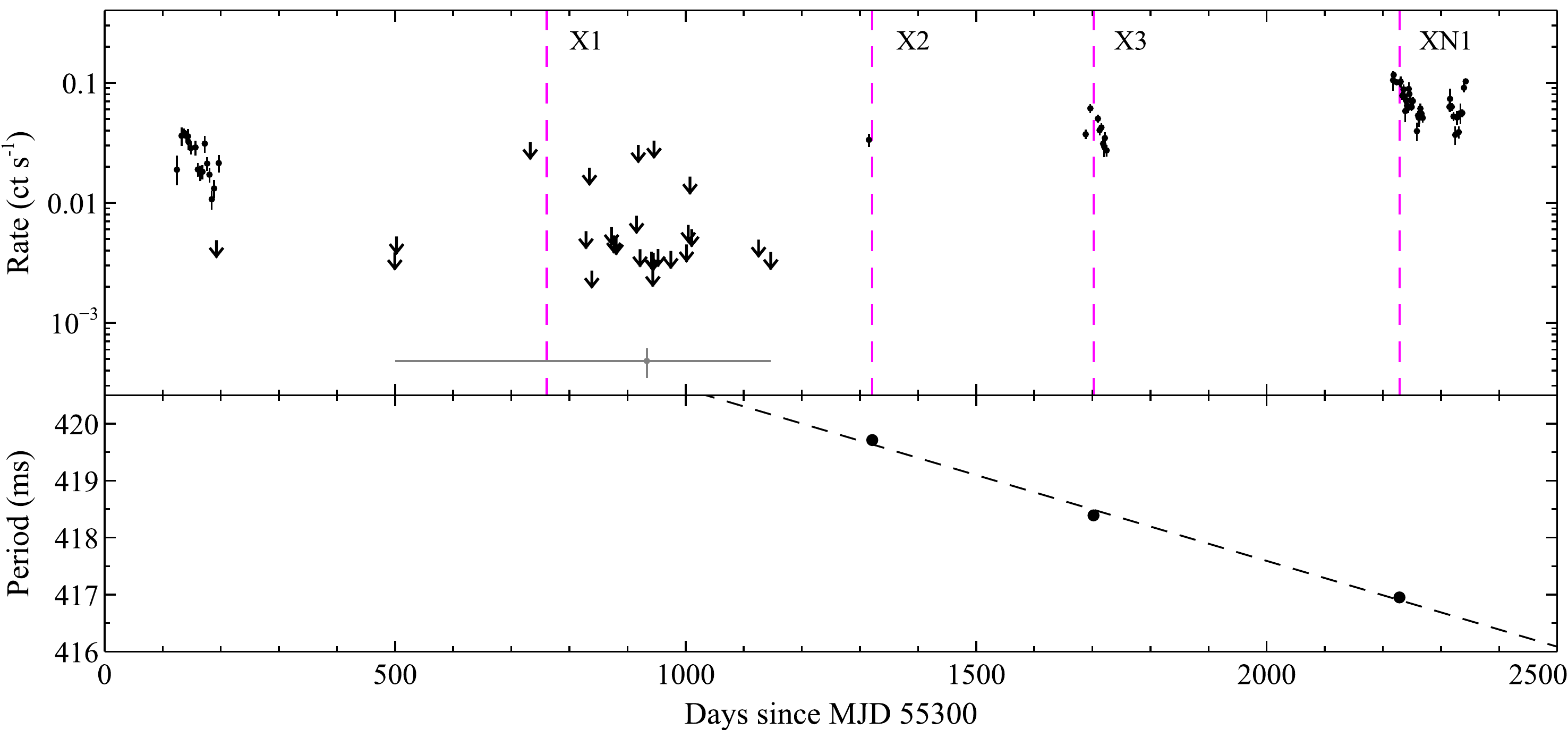}
\caption{Top: long-term 0.3--10\,keV \swift\ XRT lightcurve of P13 (1d bins), extracted with the standard \swift\ pipeline (\citealt{Evans09}). The
timing of the \xmm\ only (X1--3) and \xmm+\nustar (XN1) observations are
indicated with the dashed magenta lines. While the source spends a lot of
time in a bright state, during which it exhibits ULX luminosities
($L>2\times10^{39}$\,\ergps), during 2011 and 2012 it also exhibited an extended low flux period ($L<10^{38}$\,\ergps) indicated by the
upper limits for individual observations spanning $\rm{MJD}\approx55800-56400$ (see
\citealt{motch14a}). Stacking all the \swift observations during this period leads to
a weak detection, shown in grey, $>$2 orders of magnitude fainter than the peak flux.
Bottom: time
evolution of the pulse period for the observations bright enough for pulsations to be
detectable (X2, X3 and XN1). The dashed line shows  a linear regression through these data.
}
\label{fig_longlc}
\end{figure*}

Following the detection of a high flux with \swift\ (\citealt{SWIFT}), we triggered a
coordinated observation with \xmm\ and \nustar\ in May 2016 (total exposures
of $\approx$50 and $\approx$110\,ks, respectively). In addition to these new observations,
we also analyze  three archival \xmm\ observations. Details of these observations
 are given in Table \ref{tab_obs}, and we show them in the
context of the long-term behavior of P13 in Figure~\ref{fig_longlc}.

\begin{table*}
  \caption{Details of the X-ray observations of NGC\,7793 P13 considered in this work}
\begin{center}
\begin{tabular}{c c c c c c c}
\hline
\hline
\\[-0.1cm]
Epoch & Mission(s) & OBSID(s) & Date & $F_{\rm{0.3-10}}$ & $P$ & $\dot{P}$ \\
& & & & [$10^{-14}\,\ergcms$] & [ms] & [10$^{-10}$\,s\,s$^{-1}$] \\
\\[-0.2cm]
\hline
\hline
\\[-0.1cm]
X1 & \xmm\ & 0693760101 & 2012-05-14 & $2.0^{+1.7}_{-0.9}$ & -- & -- \\
\\[-0.2cm]
X2 & \xmm\ & 0693760401 & 2013-11-25 &  $114 \pm 3$ & $419.712 \pm 0.008$ &  $0.2^{+3.4}_{-2.8}$ \\
\\[-0.2cm]
X3 & \xmm\ & 0748390901 & 2014-12-10 & $284 \pm 5$ & $418.390 \pm 0.008$ & $-0.5^{+3.0}_{-2.5}$ \\
\\[-0.2cm]
XN1 & \xmm, \nustar\ & 0781800101, 80201010002 & 2016-05-20 & $519 \pm 7$ & $416.9513 \pm 0.0017$  & $-0.02\pm0.16$ \\
\\[-0.2cm]
\hline
\hline
\\[-0.15cm]
\end{tabular}
\label{tab_obs}
\end{center}
\end{table*}

\subsection{NuSTAR}

The \nustar data were reduced using the standard pipeline, \texttt{nupipeline}, provided
in the \nustar Data Analysis Software (v1.6.0), with standard filtering and \nustar\ CALDB v20160824.
Source products were extracted from
circular regions of radius 70$''$ for both focal plane modules (FPMA/B) using
\texttt{nuprodcuts}, with background measured from large regions on the same
detectors as P13. In addition to the standard `science' data, we maximize the
signal-to-noise (S/N) by including the `spacecraft science' data following the
procedure outlined in \cite{Walton16cyg}. This increased the total good exposure time by $\approx$10\%.
Lightcurves were extracted in the 3--78\,keV energy band with a maximal resolution of 0.1\,s.

\subsection{XMM-Newton}

The \xmm data were reduced with the \xmm Science Analysis System (v15.0.0),
following the standard prescription.\footnote{http://xmm.esac.esa.int/} Owing to its
superior time resolution of 73.4\,ms, in this work we only consider data from the EPIC-pn detector
\citep{pnref}. The raw data files were cleaned using \texttt{epchain}. Source
products were generally extracted from circular regions of radius $\approx$40$''$ (the
exception being OBSID 0693760101, during which the source was extremely faint
and a radius of 20$''$ was used) and the background measured from larger, source-free areas
on the same CCD. As recommended, we
only use single--double  patterned events.

\section{Analysis}
\label{sec:time}

\begin{figure*}
\begin{center}
\hspace{-0.25cm}
\includegraphics[width=0.45\textwidth]{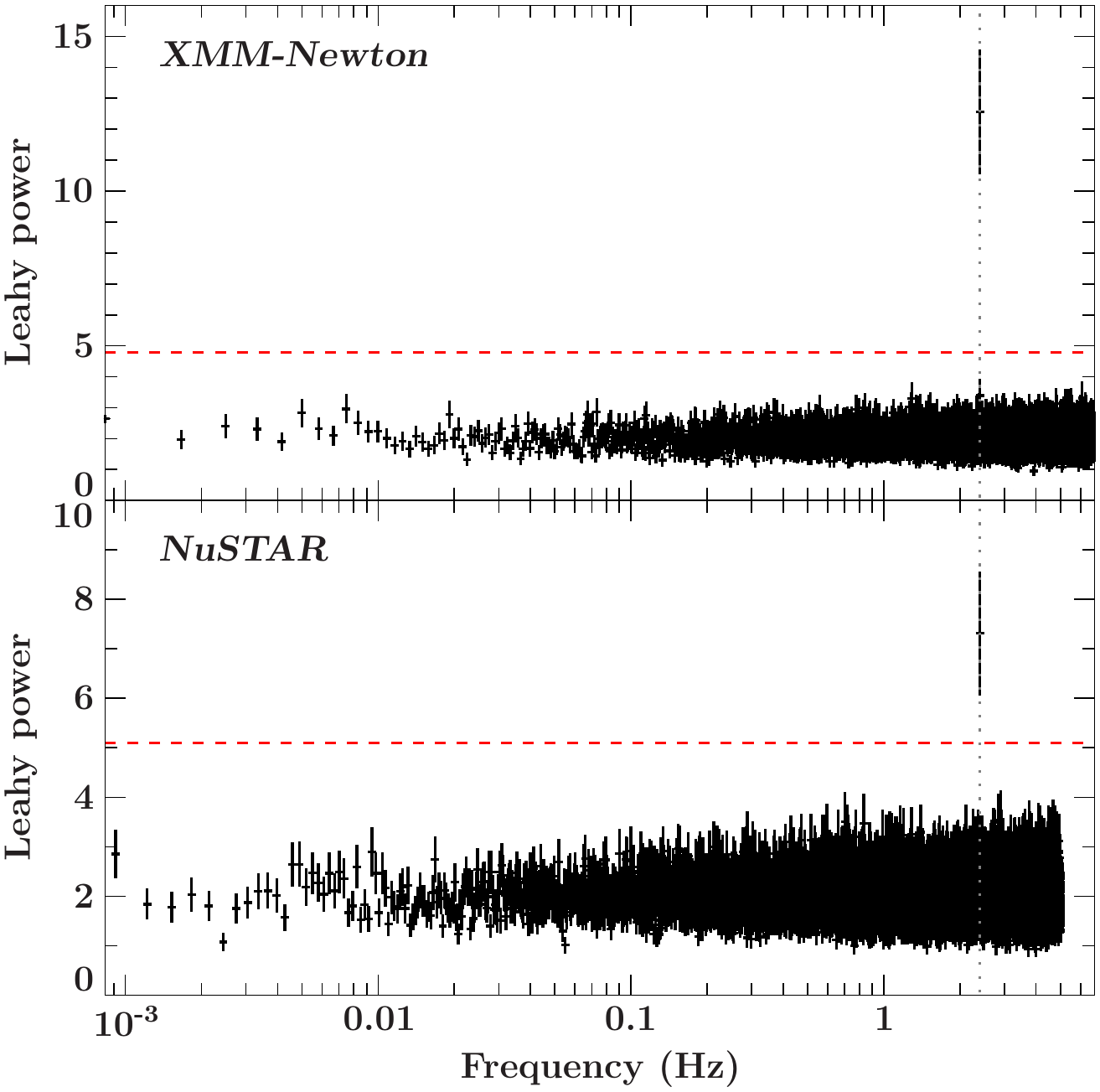}
\hspace{1.25cm}
\includegraphics[width=0.45\textwidth]{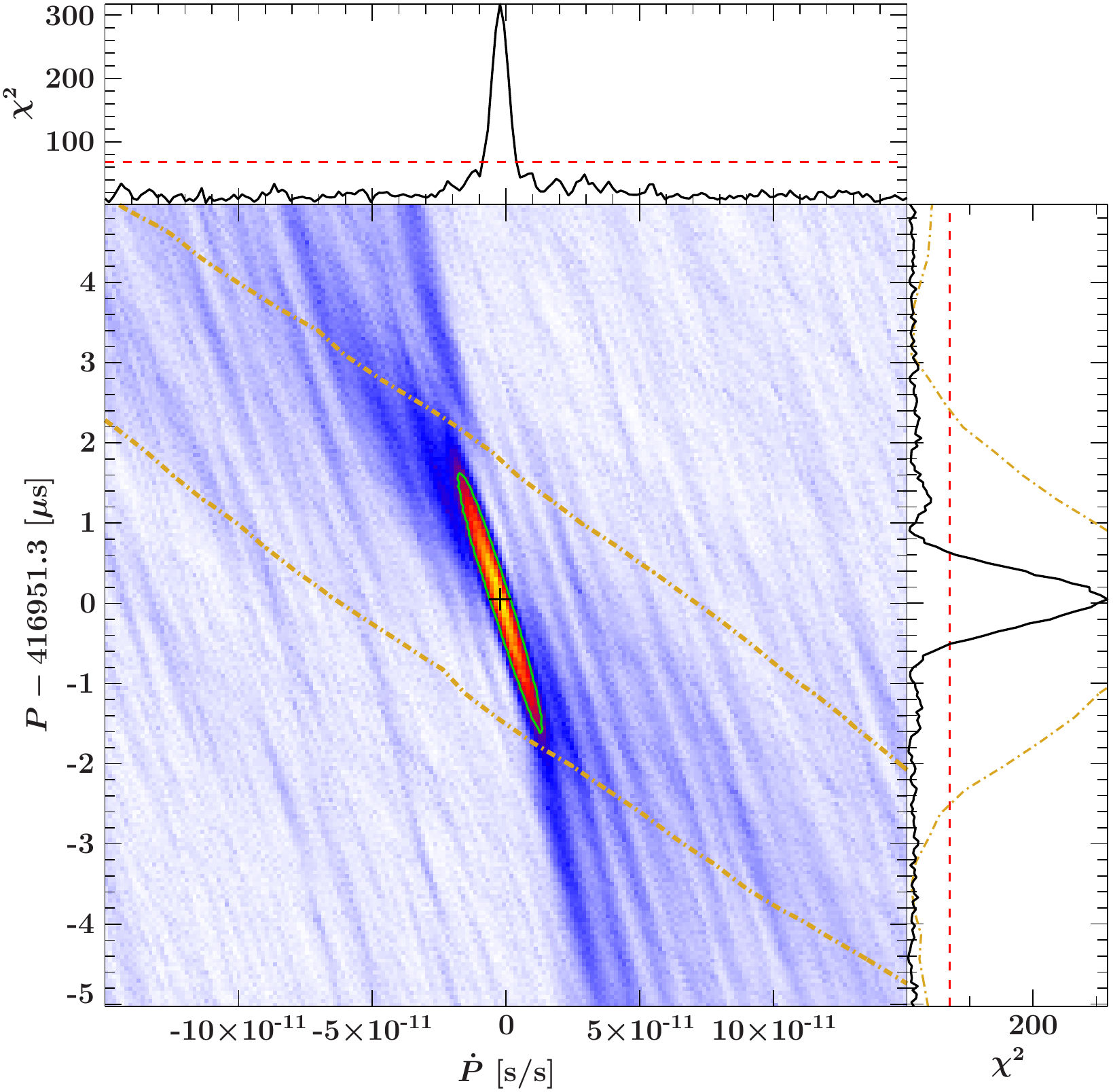}
\caption{\textit{Left:} PSD of the 2016 \xmm EPIC-pn (0.3--10\,keV, top)  and \nustar FPMA (3--78\,keV, bottom) data between 0.00124--6.812\,Hz. The
red-dashed line shows the $5\sigma$ detection limit assuming white noise
and accounting for the number of trials (8192 for \xmm and 16384 for \nustar; \citealt{vanderklis89a}). Each bin is averaged over 39 and 34 independent PSDs for \xmm and \nustar, respectively.
The gray dotted line marks 2.398\,Hz, at which a significant peak is seen in both PSDs.  \textit{Right:} result of the accelerated search of the \nustar data,
showing the $\chi^2$ values  color-coded in the $P$-$\Pdot$ plane. The green
contour line shows the estimate uncertainty following the FWHM line. The top and
right-hand panels are cuts along the respective axes through the best-fit values. The
red dashed lines in these panels indicate the $10^{-5}$ false detection probability.
The brown dotted-dashed lines shows the results for the 2016 \xmm data. }
\label{fig:psd}
\end{center}
\end{figure*}

We calculate the power spectral density (PSD) for the 2016 \xmm\ and \nustar\
lightcurves over a broad frequency range, starting at 1.24 (\xmm) and 0.3\,mHz
(\nustar) up to their respective Nyquist frequencies (6.812 and 5\,Hz, for \xmm and
\nustar). In both PSDs a strong periodic signal is evident around $\approx$2.4\,Hz, consistent between all instruments, at a significance $\gg5\,\sigma$ assuming to white noise (Figure~\ref{fig:psd}, \textit{left}).
The \nustar PSD is influenced by the variable dead time around 2.5\,ms \citep{bachetti15a}. However, the periodic signal is still significantly detected above this noise term.

To improve the period determination and search for a possible change in period
(\Pdot) we ran an accelerated epoch folding search \citep{leahy83a}. We searched
periods in a range of $\pm20\,\mu$s around the respective
peaks in the PSDs. We 
searched a range of $\pm1.5\times10^{-10}$\,\sps around $0$ for \Pdot
(Figure~\ref{fig:psd}, \textit{right}). We sampled $P$ and \Pdot on a grid of 200 values each,
and used 12 phase-bins for the pulse profiles.  

Due to their long duration ($\approx$200\,ks), the \nustar data provide the strongest constraint
on the pulse period, and we find $P=416.951\pm0.002$\,ms and
$\dot{P}=\left(-0.2\pm1.6\right)\times10^{-11}$\,\sps. The uncertainties are estimated
from the full-width, half-maximum (FWHM) contours of the $\chi^2$ landscape
(Figure~\ref{fig:psd}, \textit{right}; i.e., they account for the  degeneracy between $P$ and
\Pdot). We confirmed the values of $P$ and \Pdot through phase connection by separately folding the data of the first 50\,ks and last 50\,ks of the \nustar data and making sure no phase shift was evident.

Following this detection, we also
examined the three archival \xmm observations available. Unfortunately  the source was too
faint during the 2012 observation to perform a meaningful pulse search. However, we find significant signals in the PSDs at $\approx$2.4\,Hz in the 2013 and 2014  data. This is the only signal that is significantly detected in all calculated PSDs.

The measured values for $P$ and \Pdot are given in Table~\ref{tab_obs} and the
evolution in $P$ as a function of time is shown in Figure~\ref{fig_longlc}. While the
\Pdot measurements for the individual epochs are all consistent with zero, over the
last three years an almost linear spin-up trend is visible, with
$\dot{P}=\left(-3.486\pm0.003\right)\times10^{-11}$\,\sps. 

\begin{figure*}
\begin{center}
\hspace{-0.5cm}
\includegraphics[width=0.45\textwidth]{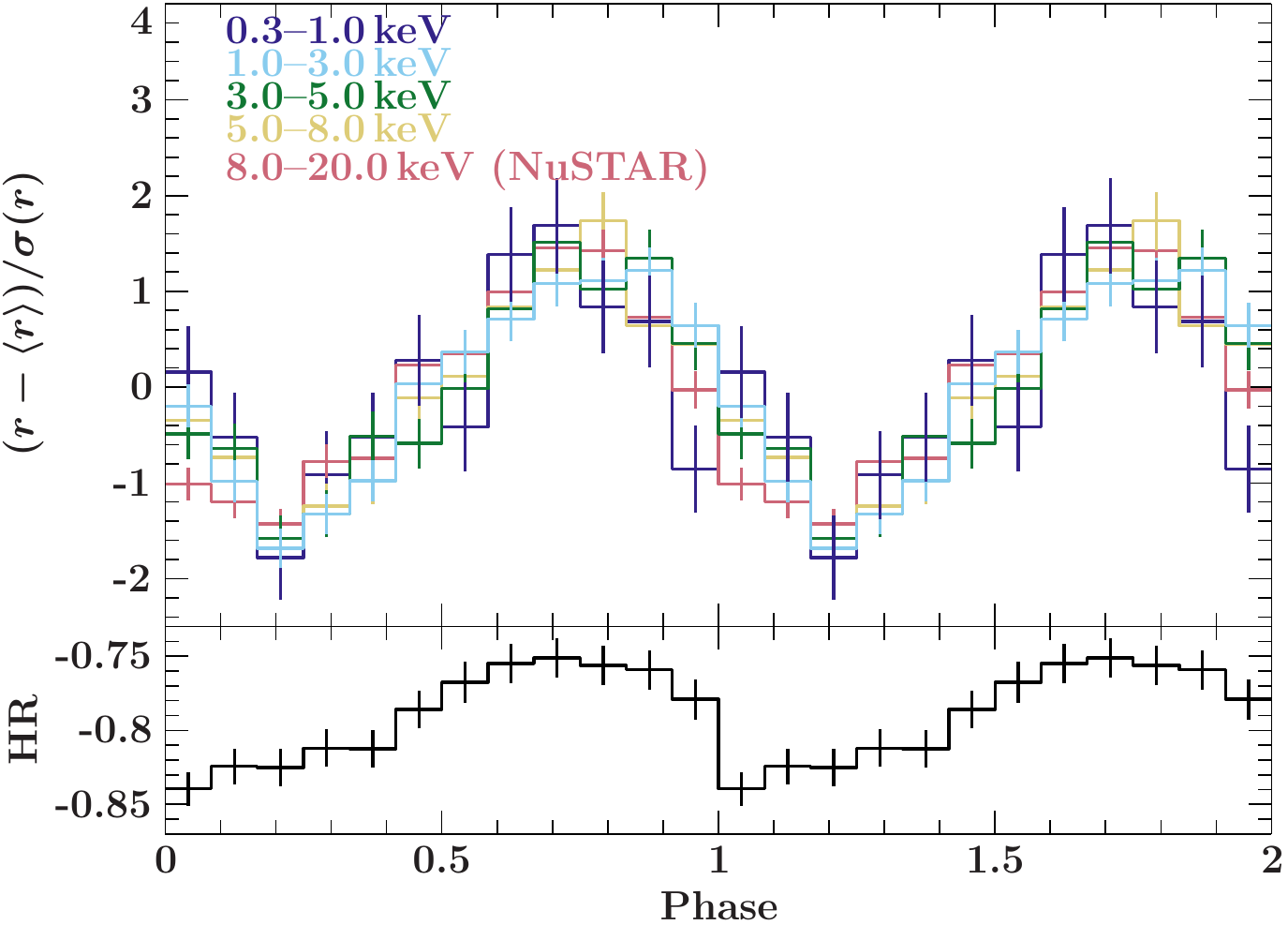}
\hspace{1.25cm}
\includegraphics[width=0.43\textwidth]{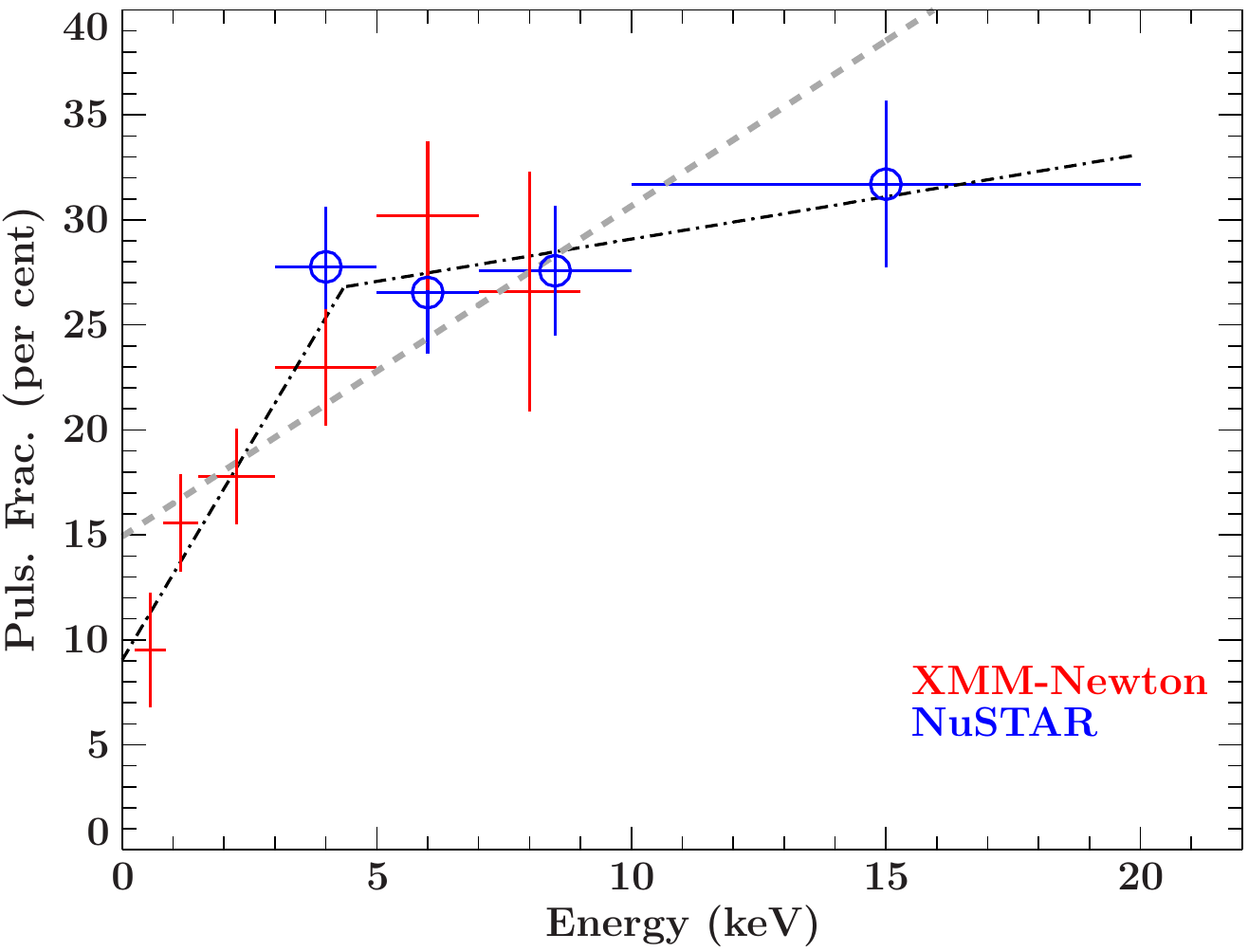}
\caption{\textit{Left, top:} Energy resolved pulse-profiles in the 0.3--1.0, 1--3, 3--5,
and 5--8\,keV bands of \xmm and the 8--20\,keV band of \nustar. The profiles
are normalized by subtracting the average rate $\left<r\right>$ and dividing by their
respective standard deviations $\sigma(r)$. 
 \textit{Bottom:} Hardness ratio between the \xmm 0.3--1\,keV (S) and the \nustar 8-20\,keV  (H) band, defined as HR=(H-S)/(H+S). \textit{Right:} Pulsed fraction as a function
of energy in \xmm (red) and \nustar (blue)  for a pulse profile with 10 phase bins. The gray dashed line gives the best-fit linear fit, while the dot-dashed black line allows for a break in slope around 4.5\,keV.}
\label{fig:profiles}
\end{center}
\end{figure*}

The 2016 data are the only available broad-band data for P13, and we use these data
to calculate energy-resolved pulse profiles between 0.3--20\,keV from the combined FPMA and FPMB events. These are shown for
five energy bands in Figure~\ref{fig:profiles}, \textit{left}.
The pulse profile is remarkably stable and sinusoidal as a  function of
energy, and agrees well between the \xmm and \nustar energy bands.

The pulsed fraction, however, changes significantly as  a function of energy (Figure~\ref{fig:profiles}, \textit{right}). The pulsed fraction PF($E$) is calculated as
\[
\text{PF}(E)=\frac{\max(p(E))-\min(p(E))}{\max(p(E))+\min(p(E))}
\]
where $p(E)$ is the energy dependent pulse profile. 
The pulsed fraction increases from 
$\approx$8\% in the 0.3--0.8\,keV band to $\approx$30\% in the 10--20\,keV band, rising
steeply before flattening out  around 5\,keV. A simple linear slope provides a statistically unacceptable description of the data; however, allowing for a break in  slope around 4.5\,keV describes the data very well (with an improvement in $\chi^2>10$ for one additional free parameter). The data can be similarly well described by a power-law  of the form $E^{\alpha}$ with an index of  $\alpha \approx 0.3$. 

The strong energy dependence of the pulsed fraction leads to significant spectral
changes over the pulse phase (Figure~\ref{fig:profiles}, \textit{left}). Here the hardness ratio between the \nustar 8--20\,keV and the \xmm
0.3--1\,keV band is shown. The spectrum is clearly softer during the low phase of the
profile and hardens during the peak. A full spectral analysis of these data will be
presented in a future work (Walton et al. \textit{in preparation}).

Standard ULX continuum models \citep[e.g,][]{Bachetti13, Walton15, Rana15} describe the data well and give an average 0.3--30\,keV flux (which dominates the bolometric emission) of $(6.7\pm0.1)\times10^{-12}$\,\ergcms. This corresponds to an apparent X-ray luminosity of $\approx$10$^{40}$\,\ergps for the NGC\,7793 distance of 3.6\,Mpc \citep{tully16a}. Scaling this to the average \swift/XRT count rate observed during 2013--2016 implies an average X-ray luminosity of $\approx7\times10^{39}$\,\ergps\ for this
period. Note that we give the 0.3--10\,keV flux  in Table \ref{tab_obs}, for comparison with the \xmm-only datasets.

\section{Discussion}
\label{sec:disc}

Our timing study of \ulx using \xmm and \nustar revealed pulsations with a  period of $\approx$418\,ms. This unambiguously identifies the compact object in the system as a neutron star. We find the pulse period in three separate epochs in 2013, 2014, and 2016. The source shows a significant average spin-up of $\approx3.5\times10^{-11}$\,\sps over the course of the past three years (Figure~\ref{fig_longlc}). 

The remarkable discovery that M82 X-2, a ULX reaching  apparent luminosities of
$L_{\rm{X}}\approx2\times10^{40}$\,\ergps, is powered by an accreting neutron star \citep{bachetti14a} naturally leads to the expectation of a larger population of ULX neutron stars
\citep{shao15a, mushtukov15a}. However, prior to
this work (and the simultaneous discovery of \citealt{israel16a}), no other examples of such systems have been confirmed
(\citealt{Doroshenko15}). The detection of X-ray pulsations from P13 implies that neutron star accretors could play a significant
role in terms of ULX demographics.

\subsection{The P13 binary system}

The identification of the compact object as a neutron star  confirms the basic conclusion of \cite{motch14a} that
P13 hosts a normal stellar-remnant.
The formal mass range for the accretor of 3--15\,\msun\ presented in that work
explicitly assumes a BH accretor. If we  allow also for a neutron star accretor, their
constraint would presumably become an upper limit of $<$15\,\msun, consistent with
the neutron star mass regime.

\citet{motch14a} interpret the 64\,d optical period as the orbital period, and  give best fit radial velocity amplitudes
between $K$=120--290\,km\,s$^{-1}$ for the compact object.
Even for a radial velocity as low as $K=120$\,km\,s$^{-1}$, this could result in a \Pdot over the long \nustar observation as large as  $2.5\times10^{-10}$\,\sps, due to the Doppler effect of the orbital motion. We rule out a \Pdot of that magnitude from \nustar timing. However, the observable \Pdot vanishes close to superior and inferior conjunction.
To reconcile the very small measured \Pdot with the orbital solution, our observation therefore has to have  taken place close to one of the conjunctions. As we do not know the exact
orbital phase of the observation, due to possible drifts with the potential $\approx$7\,yr super-orbital period
\citep{motch14a}, a more precise prediction of the expected \Pdot\ is currently not
possible.

It is still interesting to speculate whether the origin of the observed 64\,d period could be super-orbital instead of orbital as this would naturally explain the lack of observed \Pdot.
The period is similar to the $\approx$62\,d period seen from M82 X-1 or X-2 \citep{kaaret07a}.
It is most likely  associated with
X-2 (\citealt{Qiu15}), and if so must be super-orbital.
\citet{motch14a} suggest that the period in P13 is orbital, as it is also seen in the radial velocity of \heii. However, they note that the
\heii\ emission cannot arise in the stellar companion, as it is shifted in phase compared to the
photometric period. Instead, the variations in \heii\ could be related to super-orbital precession of the outer disk.
A super-orbital nature would also explain the drifts of the photometric maxima observed by \citet{motch14a}.
In this case the orbital period would be much shorter, and the lack of $\dot{P}$ in any of the individual X-ray observations would imply that we view the source
close to face-on.

As stated by \citet{motch14a}, the high luminosity rules out a purely wind-fed system.  To sustain  its luminosity  a mass accretion rate around $10^{-7}$\,\msun\,yr$^{-1}$ is required, which is comparable to the $4\pi$ mass-loss rates of B-stars \citep{vink00a}. While wind clumping can locally increase the density and influence the estimated mass loss rate \citep[see, e.g.,][]{oskinova06a}, wind-fed systems can only capture  a small fraction of that mass, even when allowing for an accretion stream. 
It is therefore likely that the companion is filling its Roche lobe \citep[radiative or atmospheric, ][]{podsiadlowski02a,fragos15a}, which distinguishes P13 from otherwise similar Galactic systems like Vela~X-1 \citep[see][and references therein]{velastat}.

\subsection{Properties of the neutron star}

In the standard disk accretion scenario as described by \citet[see also \citealt{dallosso15a}]{ghosh79a}, we can estimate the magnetic field from the change in pulse period:
\begin{equation}
-\dot{P} = 5.2\times10^{-10}{\mu_{30}}^{2/7}n(\omega_s, \mu_{30}) \left(P{L_{37}}^{3/7}\right)^2\,\text{s\,s}^{-1}
\label{eq:ppdot}
\end{equation}
assuming a neutron star mass of 1.4\,\msun and a radius of 10\,km. Here $\mu_{30}$ is  the magnetic moment in units of $10^{30}$\,G\,cm$^{-3}$ and $L_{37}$ is the luminosity in units of $10^{37}$\,erg\,s$^{-1}$. The dimensionless accretion torque  $n(\omega_s,\mu_{30})$ depends only weakly on the magnetic field, and can be approximated analytically for small values of  the fastness parameter $\omega_s$ \citep{ghosh79a}. For the relatively high spin of P13, this results in values of $n(\omega_s,\mu_{30})\ll1$. From this equation we estimate the surface magnetic field strength of P13 to be $B\approx1.5\times10^{12}$\,G.

A field of the order of $10^{12}$\,G is in the same range as expected in Galactic neutron stars, for example as measured through cyclotron resonant scattering features \citep[CRSFs; see, e.g.,][]{caballero12a,staubert14a,ks1947,cepx4}.
We note that the observed changes in $P$ seen between the different epochs, cannot be explained by  Doppler shifts due to the proposed 64\,d orbit, as this effect is orders of magnitude smaller than observed.

From Eq.~\ref{eq:ppdot} we expect a faster spin-up at higher luminosities. There is weak evidence that the average flux has been increasing over the course of the last three years, so we would therefore expect a higher spin-up in the 2016 observation than in 2013. Within each observation, the  2013 and 2014 \xmm data are not sensitive to these small values of $\dot{P}$. The long-term $\dot{P}$, however, is higher than the one measured in 2016, indicating P13 was either even brighter during the gaps in the XRT coverage, potentially reaching apparent luminosities similar to M82 X-2, or that there are additional effects that influence the accretion torque.

\subsubsection{Propeller effect}

Direct accretion can only take place if the corotation radius, $r_\text{co}$, is outside the magnetospheric radius, $r_\text{m}$. 
The magnetospheric radius for spherical accretion can be calculated as \citep{cui97a}:
\begin{equation}
r_\text{m}=2.7\times10^8{L_{37}}^{-2/7}{B_{12}}^{4/7}\,\text{cm},
\end{equation}
where $B_{12}$ is the magnetic field in units of $10^{12}$\,G.
This is also a good approximation for the magnetospheric radius of an accretion disk if the magnetic field threads the disk completely \citep{wang96a}. 
We find that  the corotation radius is about a factor of 2 larger than $r_m$ for $B_{12}=1.5$.

This relatively small difference between the radii implies that a drop in luminosity of only a factor of $\approx$10 ($r_m\propto{L_{37}}^{-2/7}$), related to natural fluctuations in the accretion rate, would therefore push P13 into the propeller regime, where $r_\text{m}>r_\text{co}$. This would then truncate the disk at large radii, dramatically reducing the accretion rate even further and suppressing the X-ray flux \citep{cui97a, tsygankov16a}, potentially explaining the very low luminosities observed before 2013.

The propeller effect has also been proposed by \citet{dallosso15a} to explain the similarly large observed changes in luminosity and transitions between low and high states seen in M82 X-2 \citep{brightman16a}.
Such variability could therefore potentially be an indicator of a ULX
pulsar and provide another avenue for identifying these systems.

\subsubsection{Intrinsic luminosity}
Assuming isotropy, we measure a luminosity of $L_{\rm{X}}\approx10^{40}$\,\ergps,
which is a factor of $\approx$50 above the spherical Eddington limit for a neutron
star. One way in which the observed flux could be reconciled with the Eddington
limit is if the emission is strongly beamed. However, this would obviously
require an overall beaming factor of $\approx$1/50. Such a  tightly collimated
beam should result in a narrow, strongly peaked pulse profile, which is clearly
at odds with the smooth, sinusoidal pulse profile we observe. The pulse profiles
of M82 X-2 and P13 are very similar in this respect.

%
This leads to a  major uncertainty in the degree to which collimation and beaming
contribute to the observed fluxes of ULX pulsars and how the magnetic field is configured.  For example for M82~X-2, different analyses provide 
strongly different estimates for the magnetic field strength, ranging over $10^9$--$10^{14}$\,G \citep{mushtukov15a,eksi15a,dallosso15a, kluzniak15a,tong15a,dallosso16a,king16a}.
So far, no model as been put forward that self-consistently reconciles the observed high luminosity, sinusoidal pulse profile, and high spin-up torque of the two known ULX pulsars.

\citet{mushtukov15a}, for example, attempt to estimate the maximum luminosity from the
accretion column as a function of B-field. For $B_{12}=1.5$, their calculations imply a maximum luminosity of
$\approx5\times10^{38}\,\ergps$, which is still a factor of $\approx$20 below our measurement.
The required beaming in this mode would still likely be at
odds with the observed pulse profile.
A better understanding of collimation and beaming in super-Eddington neutron stars is clearly necessary to link intrinsic and observed, isotropic luminosities in these systems, and more robustly constrain their magnetic fields.

\section{Conclusions}
We detected coherent X-ray pulsations from the ULX \ulx, making this only the
second confirmed ULX pulsar after M82~X-2 \citep{bachetti14a}. Its properties seem
to be in line with a high luminosity extension of known Galactic neutron star binaries.
Between observations in  2013 and 2016  we see a significant spin-up
from which we estimate the magnetic field strength to be $\approx1.5\times10^{12}$\,G, typical of Galactic
systems. Continued monitoring of the pulse period evolution of this remarkable
source will be of particular value and help us understand if high variability is a tell-tale sign of super-Eddington neutron stars.

\section*{ACKNOWLEDGEMENTS}
We would like the thank the referee for the helpful comments.
MJM
acknowledges support from an STFC Ernest Rutherford fellowship, ACF
acknowledges support from ERC Advanced Grant 340442, and DB acknowledges
financial support from the French Space Agency (CNES).
This research has made
use of data obtained with \nustar, a project led by Caltech, funded by NASA and
managed by NASA/JPL, and has utilized the \texttt{nustardas} software package, jointly
developed by the ASDC (Italy) and Caltech (USA). This research has also made
use of data obtained with \xmm, an ESA science mission with instruments and 
contributions directly funded by ESA Member States. This work made
use of data supplied by the UK Swift Science Data Centre at
the University of Leicester, and also made use of the XRT
Data Analysis Software (XRTDAS) developed under the responsibility
of the ASI Science Data Center (ASDC), Italy.
This research has made use of a collection of ISIS functions (ISISscripts) provided by ECAP/Remeis observatory and MIT (\url{http://www.sternwarte.uni-erlangen.de/isis/})

\textit{Facilities:} \facility{NuSTAR}, \facility{XMM}, \facility{Swift}

%

\end{document}